# Booster High-level RF Frequency Tracking Improvement Via the Bias-Curve Optimization

Xi Yang

*Fermi National Accelerator Laboratory*

Box 500, Batavia IL 60510

**Abstract**

It is important to improve the frequency tracking between the RF drive and the cavity field for the purpose of reducing longitudinal phase oscillations and increasing the effective accelerating voltage. And this is especially beneficial for Booster running at higher intensity with smaller beam emittance. Optimizing the bias supply current curve for the ferrite tuners can reduce the phase error between the RF drive and the cavity field, and also improve the HLRF-to-LLRF frequency tracking efficiency.

**Introduction**

The Booster accelerates protons ($H^+$) from 400 MeV to 8 GeV after the $H^-$ beam injected from the Linac is stripped to protons. The RF frequency changes from 37.9 MHz to 52.9 MHz to accelerate the beam in a Booster cycle. Each Booster RF cavity contains three ferrite tuners (FT), which are connected to the accelerating cavity and are part of the resonating structure.[1] The cavity resonant frequency is adjusted by changing the bias supply current for the purpose of varying the inductance of FT.

The relationship between the bias current and the cavity resonant frequency has been measured and applied as the bias curve (BC) for a Booster cycle. However, there are some factors, such as the hysteresis of the ferrite, Booster rep rate, temperature, and beam loading, etc., which could change the relationship between the bias current and the cavity resonance frequency and make the required bias current for a cavity resonance frequency different from the measurement. Also, since for a required rate of acceleration, the low-level RF system (LLRF) is continuously adjusting the RF drive to the cavity field to maintain the correct phase relation between circulating beam bunches and the accelerating gap voltage, the cavity resonance frequency varies slightly from cycle to cycle.[2] Since the required cavity resonance frequency couldn't be controlled only by a predetermined BC, a feedback loop (bias running closed loop) had been installed for each Booster RF station to minimize the phase error between the RF drive and the cavity field.[3]

With the continuously increased requirement for Booster to deliver protons with higher intensity and smaller emittance, it is important for us to improve the frequency tracking between the RF drive and the cavity field for the purpose of reducing phase oscillations longitudinally and increasing the effective accelerating voltage. Besides, the feedback loop only can correct the phase error between the RF drive and the cavity field after the error appears. Also, the correction time is limited by the time constant of the bias-tuning system and is about half a millisecond. A procedure, which can be used to generate the optimal BC based upon the phase error signal from the pre-Booster cycle and the cavity Q factor, has been developed to improve the HLRF-to-LLRF frequency tracking.

## Method

The bias supply current for each RF station is the sum of three inputs, the BC signal, the phase error signal and the DAC signal. The DAC signal provides an offset to the bias supply current in the beginning of a cycle and varies from station to station. Here, we set it to zero. The BC signal is obtained from Booster console program B30 with a calibration constant of 250 A/V, as shown in Fig. 1(a). The phase error signal is generated by the analog device AD532 whose transfer function is [4]

$$I_{pes}(t) = \left((X_1(t) - X_2(t)) \cdot (Y_1(t) - Y_2(t))\right)/10. \qquad 1$$

Here, $t$ is the time in a Booster cycle, $I_{pes}(t)$ is the phase error signal, and $X_1(t)$ is the phase error from the phase detector. The phase detector is continuously comparing the phase difference between the cavity field and the RF drive and outputting it as the phase error in unit of volt with a calibration constant of 1 V/9°. $Y_1(t)$ is the BC signal in unit of ampere, $X_2(t)$ and $Y_2(t)$ are zero due to the grounding.

The phase error taken at a MiniBooNE event ($1D) from station #16 is shown in Fig. 1(b). The phase error signal (PES), which is the product of Fig. 1(a) and Fig. 1(b) and then divided by 10, is shown in Fig. 1(c). The total bias supply current, which is the sum of Fig. 1(a) and Fig. 1(c), is shown in Fig. 1(d). It is important for us to notice that when station #16 was running at the bias current curve shown in Fig. 1(d) the PES still appeared as the curve shown in Fig. 1(c).

The frequency difference ($\Delta f$) between the cavity field and the RF drive can be calculated using the parallel-RLC circuit model,[5] as shown in eq.2, once the phase error from the phase detector is known.

$$\frac{1}{Z(\omega)} = \frac{1}{R_s} + \frac{1}{j\omega L} + j\omega C, \qquad 1^{st}$$

$$\text{Re}(Z(\omega)) = \frac{\omega^2 L^2 R_s}{R_s^2(1-\omega^2 LC)^2 + \omega^2 L^2}, \qquad 2^{nd}$$

$$\text{Im}(Z(\omega)) = \frac{j\omega L R_s^2(1-\omega^2 LC)}{R_s^2(1-\omega^2 LC)^2 + \omega^2 L^2}, \qquad 3^{nd}$$

$$\tan(\Delta\phi(\omega)) = \frac{\text{Im}(Z(\omega))}{\text{Re}(Z(\omega))} = \frac{jR_s(1-\omega^2 LC)}{\omega L}, \qquad 4^{th}$$

$$\omega = \omega_0 + \Delta\omega, \qquad 5^{th}$$

$$\Delta f = f_0 - f_d = \frac{\Delta\omega}{2\pi} \approx -(\tan(\Delta\phi(\omega))) \cdot \left(\frac{\omega_0}{4\pi \cdot Q_0}\right), \qquad 6^{th}$$

$$\Delta\phi(\omega) = (V_{pes}) \cdot 9 \cdot \frac{\pi}{180}, \qquad 7^{th}$$

$$\omega_0 = 2\pi \cdot f_0 = \frac{1}{\sqrt{LC}}. \qquad 8^{th}$$

Here, $Z(\omega)$ is the impedance of the parallel-RLC circuit at the angular frequency $\omega=2\pi f$, $R_s$ is the shunt impedance, $L$ is the inductance, $C$ is the capacitance, $f_0$ is the cavity resonance frequency, $Q_0$ is the cavity quality factor at the resonance frequency, $f_d$ is the RF drive frequency, $V_{pes}$ is the phase error between the cavity field and the RF drive in unit of volt, and $\Delta\phi(\omega)$ is the phase angle of $Z(\omega)$ in unit of radian. Small angle approximation and Taylor expansion of $\omega=\omega_0+\Delta\omega$ to the 1$^{st}$ order of $\Delta\omega$ are used in the calculation of $\Delta f$ from $\Delta\phi(\omega)$, $Q_0$, and $\omega_0$, as shown in the 6$^{th}$ equation in eq.2.

The cavity $Q$ vs. time $t$ in a Booster cycle is shown in Fig. 2(a).[2] The RF drive frequency ($f_d$) for each Booster cycle is calculated from the magnet ramp,[6] and is shown in Fig. 2(b). From the past experience, the difference between the HLRF frequency and the LLRF drive frequency couldn't be bigger than half percent; otherwise part of the beam will fall off. We can use $2\pi f_d$ as $\omega_0$ in the 6$^{th}$ equation of eq.2 and get the frequency error $\Delta f$ for the $1D event, as shown in Fig. 2(c). The cavity resonance frequency is obtained using $f_d + \Delta f$ (the sum of Fig. 2(b) and Fig. 2(c)), and the total bias current vs. the cavity resonance frequency is shown in Fig. 2(d). The relationship between the bias current ($I$) and the cavity resonance frequency is obtained by the exponential fit of the curve in Fig. 2(d), as shown in eq.3(a).

$$I_{bc}(f) = a \cdot \exp(b \cdot f) = 0.7768 \cdot \exp(0.1505 \cdot f). \qquad 3(a)$$

Here, the unit of $f$ is MHz. We obtain eq.3(b) by differentiating eq.3(a).

$$\Delta I_{bc}(f) = a \cdot b \cdot \exp(b \cdot f) \cdot \Delta f \approx 0.1169 \cdot \exp(0.1505 \cdot f) \cdot \Delta f \qquad 3(b)$$

The amount of bias current, which should be adjusted in order to eliminate the phase error, should be $-\Delta I_{bc}$, and the result is shown in Fig. 2(e). The optimal bias current should be the sum between the total bias current and $-\Delta I_{bc}$, as shown in Fig. 2(f). However, $\Delta I_{bc}$ is less than 0.3% of the total bias current, and this can be understood by the calculation shown in eq.4, which is obtained through dividing eq.3(b) by eq.3(a).

$$\frac{\Delta I_{bc}(f)}{I_{bc}(f)} = b \cdot \Delta f \leq b \cdot |\Delta f|_{max} \approx 0.23\%. \qquad 4$$

**Comment**

The difference between the bias curve shown in Fig. 1(d) and the one shown in Fig. 2(f) is less than 0.3%, and more efforts are required for the optimal bias current curve in Fig. 2(f) than those for the total bias current curve in Fig. 1(d). Also some other fluctuations in Booster from cycle to cycle can be comparable or larger than 0.3%. So it is helpful for us to modify the control system of the bias supply current and use the total bias current curve, which is obtained from the pre-Booster cycle, as the bias current curve for the present cycle and keep on updating the bias current curve frequently, at least from day to day. Whenever it is possible, tuning the bias current curve to minimize the average phase error signal from all the stations will be helpful. And the reason why we use the average phase error signal for tuning is because all the stations are running the same BC. Using a separated bias current curve for each RF station can remove the phase error from the station-to-station difference and improve the HLRF-to-LLRF frequency tracking further.

## Acknowledgements

The author gives thanks for the tremendous amount of help from the HLRF group in Fermilab, especially to Rene Padilla, John Reid, Tim Berenc and Bob Scala, who provided the author with the technical information and guidance for this work.

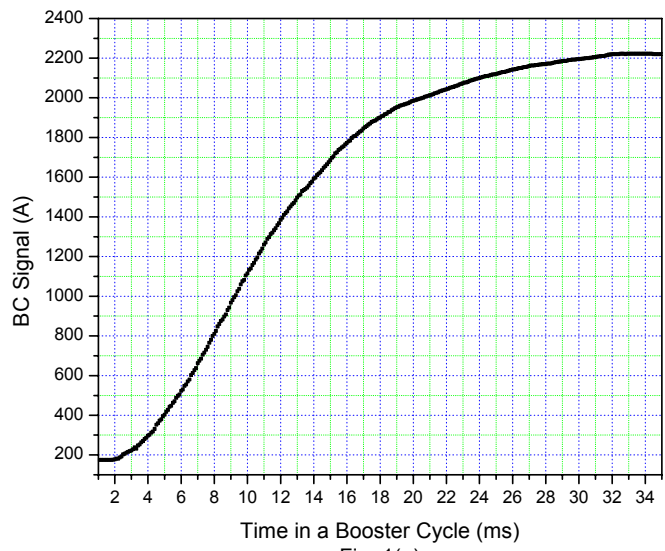

Fig. 1(a)

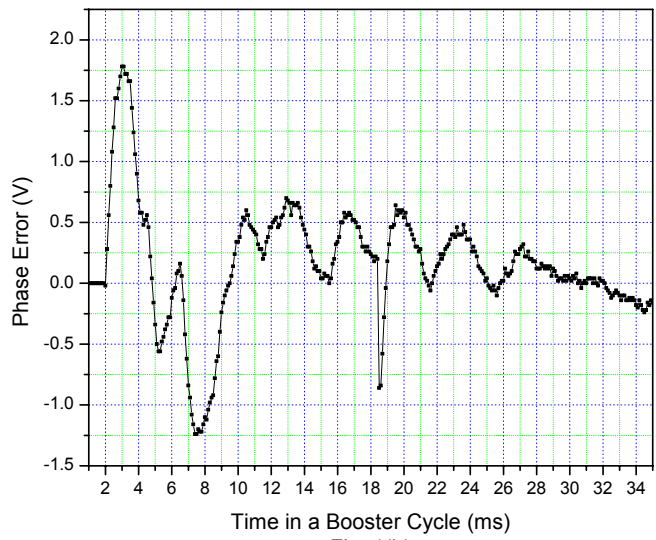

Fig. 1(b)

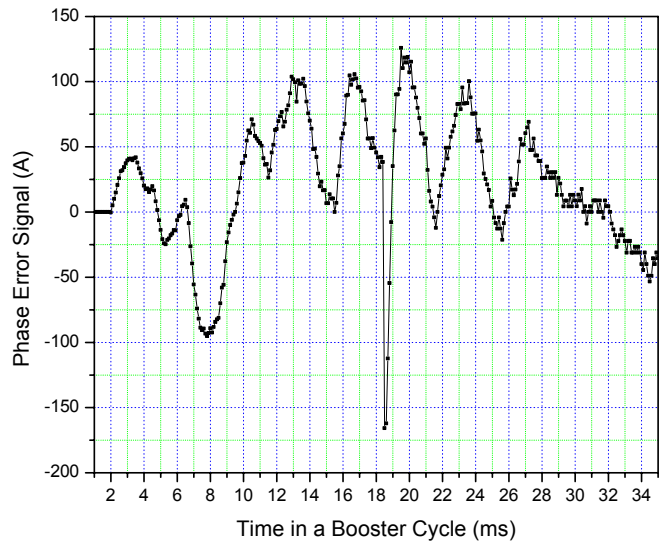

Fig. 1(c)

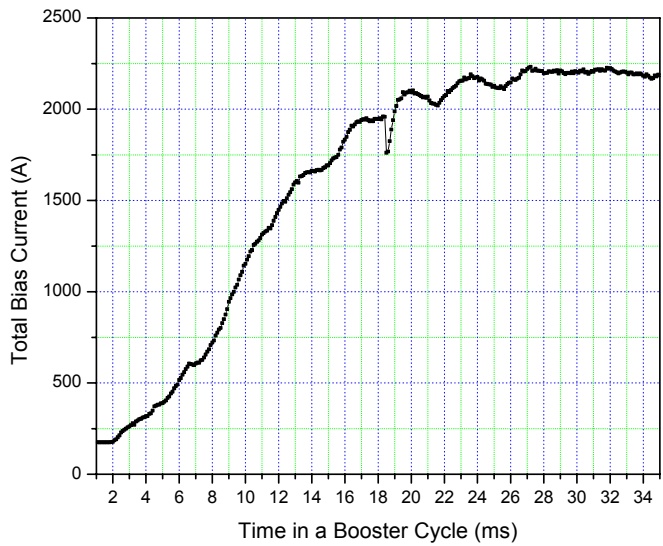

Fig. 1(d)

Fig. 1(a) the bias curve obtained from Booster console program B30 with a calibration of 250 A/V.

Fig. 1(b) the phase error taken at a MiniBooNE event ($1D) from station #16.

Fig. 1(c) the phase error signal, which is the product of Fig. 1(a) and Fig. 1(b) and then divided by 10.

Fig. 1(d) the total bias supply current, which is the sum of Fig. 1(a) and Fig. 1(c).

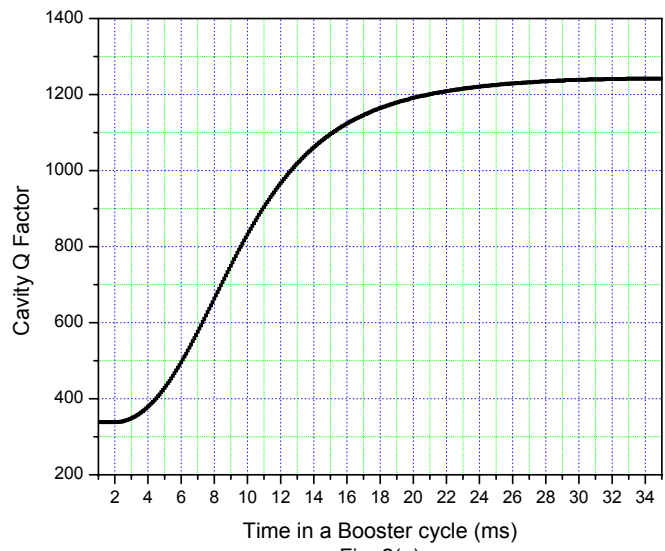

Fig. 2(a)

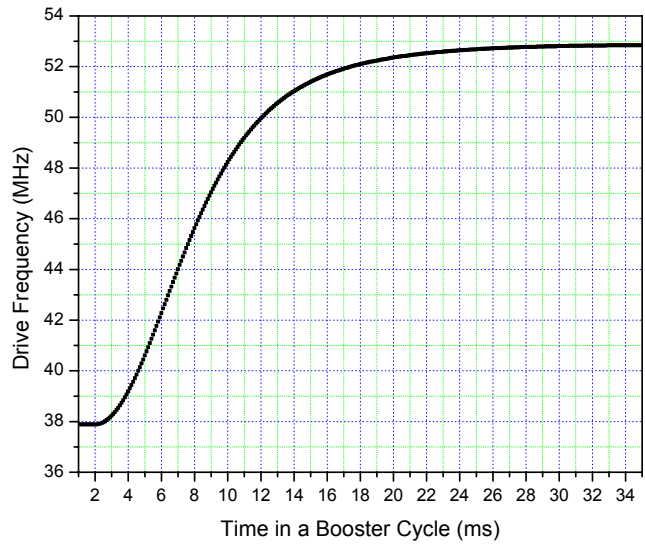

Fig. 2(b)

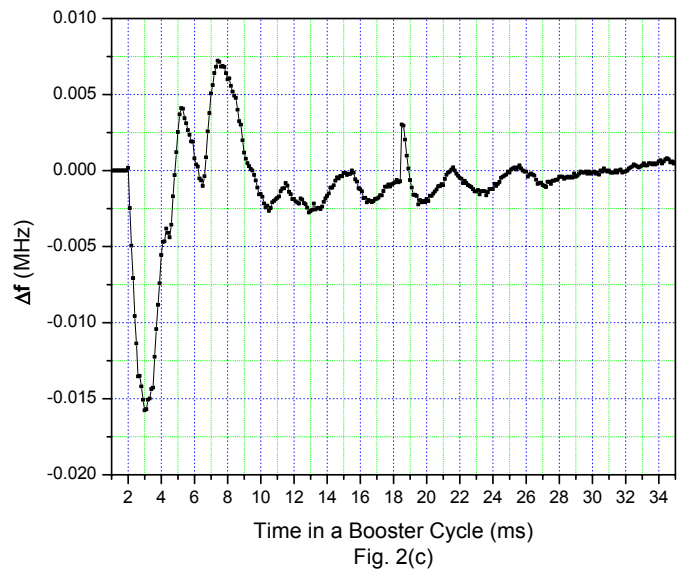

Fig. 2(c)

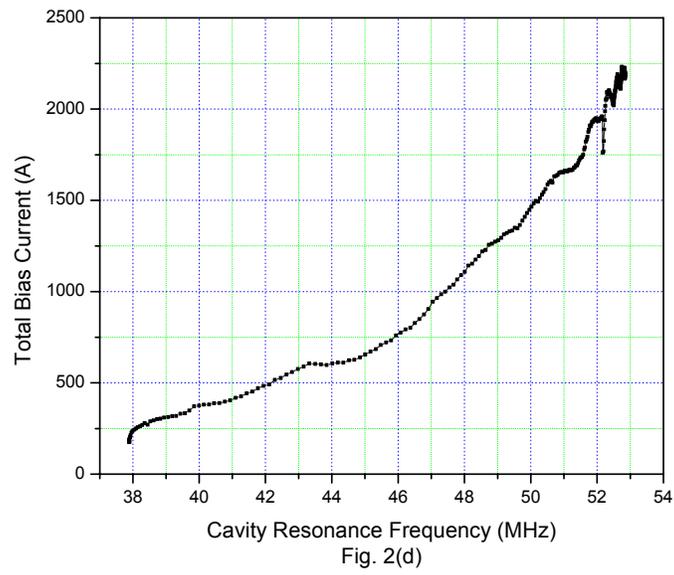

Fig. 2(d)

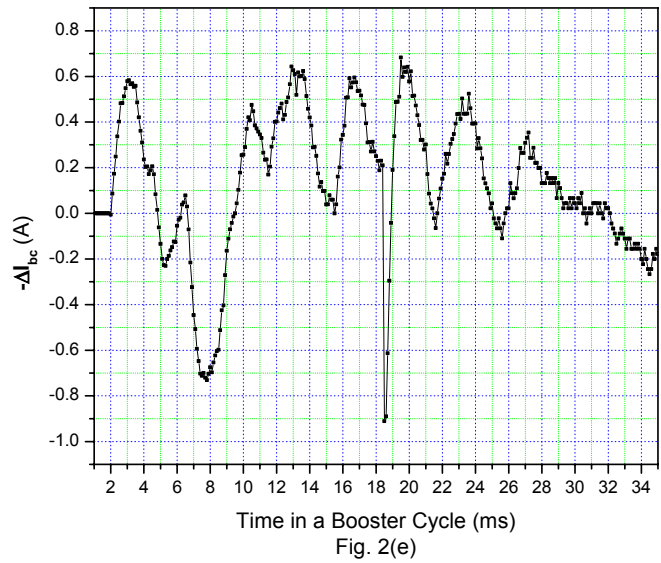

Fig. 2(e)

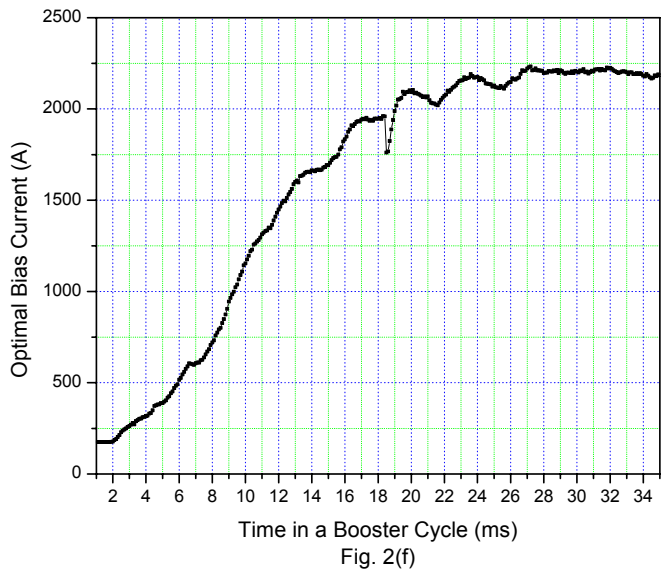

Fig. 2(f)

Fig. 2(a) the cavity $Q$ *vs*. time $t$ in a Booster cycle.

Fig. 2(b) the RF drive frequency ($f_d$) for a Booster cycle.

Fig. 2(c) the frequency error ($\Delta f$) of station #16 for the $1D event shown in Fig. 1(b).

Fig. 2(d) the total bias current *vs*. the cavity resonance frequency.

Fig. 2(e) the amount of bias current, which should be adjusted in order to eliminate the phase error shown in Fig. 1(b).

Fig. 2(f) the optimal bias current.